\documentclass{goose-article}

\header{%
  T.W.J.~de~Geus, M.~Cottura, B.~Appolaire, R.H.J.~Peerlings, M.G.D.~Geers\\%
  Mechanics of Materials, 97:199--211, 2016, %
  \doi{10.1016/j.mechmat.2016.02.006}%
}

\title{Fracture initiation in multi-phase materials: a systematic three-dimensional approach using a FFT-based solver}

\author[1,2]{T.W.J.~de~Geus$^*$}
\author[1,2]{M.~Cottura}
\author[3]{B.~Appolaire}
\author[2]{R.H.J.~Peerlings}
\author[2]{M.G.D.~Geers}

\affil[1]{
  Materials innovation institute (M2i),
  P.O.~Box~5008, 2600~GA~Delft, The~Netherlands
}
\affil[2]{
  Department of Mechanical Engineering, Eindhoven University of Technology,\nl
  P.O.~Box~513, 5600~MB~Eindhoven, The~Netherlands
}
\affil[3]{
  Laboratoire d'Etude des Microstructures, CNRS/Onera,
  BP72, 92322 Ch\^{a}tillon Cedex, France
}

\contact{%
  $^*$Corresponding author:
  \href{mailto:t.w.j.d.geus@tue.nl}{t.w.j.d.geus@tue.nl} %
  \hspace{1mm}--\hspace{1mm} %
  \href{mailto:tom@geus.me}{tom@geus.me} %
  \hspace{1mm}--\hspace{1mm} %
  \href{http://www.geus.me}{www.geus.me}%
}

\begin{document}
% %%%%%%%%%%%%%%%%%%%%%%%%%%%%%%%%%%%%%%%%%%%%%%%%%%%%%%%%%%%%%%%%%%%%%%%%%%%%%%

\maketitle

% ------------------------------------------------------------------------------
\begin{abstract}
% ------------------------------------------------------------------------------

This paper studies a two-phase material with a microstructure composed of a hard brittle reinforcement phase embedded in a soft ductile matrix. It addresses the full three-dimensional nature of the microstructure and macroscopic deformation. A large ensemble of periodic microstructures is used, whereby the individual grains of the two phases are modeled using equi-sized cubes. A particular solution strategy relying on the Fast Fourier Transform is adopted, which has a high computational efficiency both in terms of speed and memory footprint, thus enabling a statistically meaningful analysis. This solution method naturally accompanies the regular microstructural model, as the Fast Fourier Transform relies on a regular grid.

Using the many considered microstructures as an ensemble, the average arrangement of phases around fracture initiation sites is objectively identified by the correlation between microstructure and fracture initiation -- in three dimensions. The results show that fracture initiates where regions of the hard phase are interrupted by bands of the soft phase that are aligned with the direction of maximum shear. In such regions, the hard phase is arranged such that the area of the phase boundary perpendicular to the principal strain direction is maximum, leading to high hydrostatic tensile stresses, while not interrupting the shear bands that form in the soft phase. The local incompatibility that is present around the shear bands is responsible for a high plastic strain. By comparing the response to a two-dimensional microstructure it is observed that the response is qualitatively similar (both macroscopically and microscopically). One important difference is that the local strain partitioning between the two phases is over-predicted by the two-dimensional microstructure, leading to an overestimation of damage.

% ------------------------------------------------------------------------------
\end{abstract}
% ------------------------------------------------------------------------------


% ==============================================================================
\section{Introduction}
% ==============================================================================

Multi-phase materials are used in a wide variety of applications because they often encompass good engineering properties: electrical, thermal, mechanical, etc. By combining two or more distinct phases at the level of the microstructure, a combination of properties is obtained that cannot always be achieved otherwise. This paper focuses on the class of two-phase materials that display both strength and ductility using a microstructure in which a soft and ductile matrix is reinforced using a hard and brittle phase.

The elasto-plastic response of such materials is reasonably well understood. Simple rules of mixture yield satisfactory predictions for many purposes \cite{Koo1980, Davies1978}, but also mean-field \cite{Maire1997, Gonzalez2000} and unit cell models have been applied \cite{Brockenbrough1990, Brockenbrough1995, Povirk1995, Schroder2010}. Many of them have been extended to incorporate microstructural damage to predict the loss of material integrity observed at the macro-scale. However, in most cases, the role of the microstructure, and in particular of the phase distribution, is only partially considered and not fully understood.

Different researchers have performed studies that aimed at characterizing the role of the spatial distribution of the phases on the initiation and growth of fracture at the level of individual grains \cite{Kumar2006, LLorca2004, Segurado2006, DeGeus2015a}. In these papers it was found that damage preferentially nucleated in or in-between reinforcement particles that are neighboring along the tensile axis -- see also the experimental work by Lewandowski et al.\ \cite{Lewandowski1989}. A positive hydrostatic stress was one of the configurational characteristics driving damage nucleation and growth.

Recently, De Geus et al.\ \cite{DeGeus2015a} have demonstrated that the role of the microstructure can be rationalized by considering the average phase distribution around damage sites through a conditional spatial average. The strength of this approach is that the results need little interpretation as the correlation between morphology and fracture initiation is quantified. Besides confirming the role of neighboring reinforcement particles in the tensile direction, the critical microstructure features a pattern composed of matrix phase directly adjacent to the damage site in the direction perpendicular to the tensile direction, and in bands in the directions of maximum shear. Later, it was shown that this phase distribution is also critical for fracture initiation by cleavage in the reinforcement phase \cite{DeGeus2015b}.

Most published models are restricted to two spatial dimensions. Likewise experimental observations generally are restricted to cross-sections. In spite of the valuable insights gained, the impact of the geometrical approximation remains unclear. Some studies were performed in three dimensions \cite{LLorca2004,Segurado2006,Ramazani2013}, but a systematic analysis including many statistical variations is often unfeasible. Ramazani et al.\ \cite{Ramazani2013} observed a clear difference in the macroscopic mechanical properties predicted by two- and three-dimensional micromechanical models. For a different physical process, Xu et al.\ \cite{Xu2013} compared the accuracy of two-dimensional representative volume elements (RVEs) to full three-dimensional RVEs in the context of humidity induced electrochemical degradation of fuel cells with a Nickel -- Yttria-stabilized Zirconia microstructure. They concluded that the yield stresses are of the same order of magnitude in the two-dimensional model as in the three-dimensional model, but the former cannot capture the thermoelectrical properties accurately as its percolation is inherently limited. These distinct mechanisms may also have a strong influence on fracture initiation for the class of materials considered here, as the plastic strain localizes in a path of soft phase that is connected in the direction of shear, prior to fracture initiation.

A systematic analysis of the local mechanical response of a three-dimensional microstructure, with statistical relevance, is still beyond the capabilities of conventional computational solid mechanics, based on the finite elements method. Different methods have been developed that enable such an analysis. Among them, the Voronoi tesselation finite elements method of Moorthy and Ghosh \cite{Moorthy1998a} is worth mentioning. This method employs a microstructure that is generated using a Voronoi tessellation. Each Voronoi cell forms an element for which the response is calculated using an assumed stress distribution. Although the resulting number of degrees of freedom is limited, the extension to three dimensions is still non-trivial \cite{Ghosh2004}.

Besides the different strategies for improving the efficiency of the calculations based on the finite elements method, full-field calculations based on the Fast Fourier Transform (FFT), as first proposed by Moulinec and Suquet \cite{Moulinec1998}, have proved to be a serious alternative for periodic inhomogeneous media when the number of degrees of freedom becomes significant, as it is the case in 3-D. This method takes advantage of the fact that equivalent linear elastic problems can be formulated and recast into integral forms, where the relevant Green functions are convolved with some appropriate source fields. The convolution product is expressed as a simple product in Fourier space and is computationally evaluated by means of an efficient FFT-library. Even though the application of the Fast Fourier Transform restricts the discretization to a regular grid, this method has the great advantage that the computational costs, both in terms of CPU time and memory footprint, scale approximately linearly with the number of grid points. The method can thus handle large systems and appears to be the best suited for 3-D microstructures.

The current paper aims to extend the analysis by De Geus et al.\ \cite{DeGeus2015a} to a three-dimensional setting. By comparing different three-dimensional deformation paths for a large number of three-dimensional microstructures, the role of the microstructural morphology on fracture initiation is systematically identified. The results are compared to those obtained from two-dimensional microstructures, complementing the analyses of fracture initiation mechanisms but also allowing to identify the limitations of such two-dimensional models. This is achieved by means of the algorithm proposed by Moulinec and Suquet \cite{Moulinec1998} to solve mechanical equilibrium in heterogeneous materials. Besides its low memory footprint, the simplifying but reasonable assumption that the elastic moduli are homogeneous renders the method incomparably fast and the most suitable for our statistical analysis.

This paper is structured as follows. The microstructural model and the numerical implementation are discussed in Sections~\ref{sec:model} and~\ref{sec:numerical} respectively. In Section~\ref{sec:res:microstructure}, the two- and three-dimensional microstructures are compared in terms of macroscopic and microscopic elasto-plastic response and fracture initiation. For the three-dimensional microstructures, the effects of different deformation paths are investigated in Section~\ref{sec:res:deformation}. The paper ends with several concluding remarks in Section~\ref{sec:conclusion}.

% ==============================================================================
\section*{Nomenclature}
% ==============================================================================

\begin{tabular}{llll}
  $\bm{A}$
  & second order tensor
  \\
  $\mathbb{A}$
  & fourth order tensor
  \\
  $\mathbb{C} \, = \bm{A} \otimes \bm{B}$
  & dyadic tensor product ($C_{ijkl} = A_{ij} B_{kl}$)
  \\
  $\bm{C} = \bm{A} \cdot \bm{B} $
  & single tensor contraction ($C_{ik} = A_{ij} B_{jk}$)
  \\
  $c \;\, = \bm{A} : \bm{B}$
  & double tensor contraction ($c = A_{ij} B_{ji}$)
  \\
  $\langle a \rangle$
  & ensemble average
  \\
  $\bar{a}$
  & volume (unit cell) average
  \\
\end{tabular}

% ==============================================================================
\section{Model}
\label{sec:model}
% ==============================================================================

% ==============================================================================
\subsection{Microstructure}
% ==============================================================================

\begin{figure}[tph]
  \centering
  \includegraphics[width=0.8\linewidth]{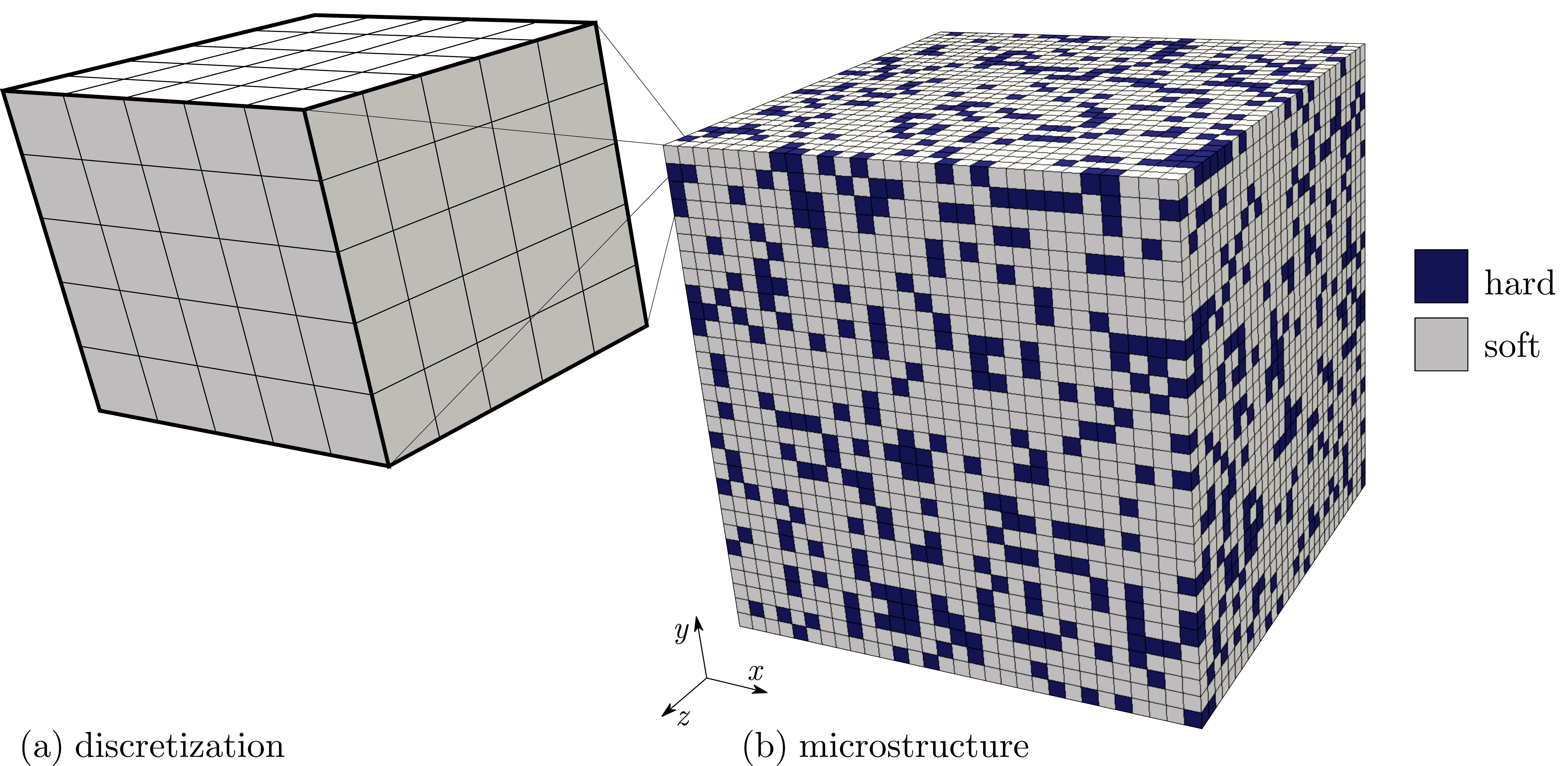}
  \caption{Typical unit cell from the ensemble of $100$ unit cells representing the two-phase material (b). The numerical discretization of a single grain; each grain is discretized in the same way (a).}
  \label{fig:microstructure}
\end{figure}

The microstructure is represented by an ensemble of $100$ three-dimensional periodic unit cells. Each unit cell comprises $30 \times 30 \times 30$ cubes arranged in a regular grid, see for a typical example Figure~\ref{fig:microstructure}(b). These cubes represent the individual grains or particles of the two phases considered -- a soft and a hard phase with respectively low and high yield stresses. They are referred to as \emph{grains} below, to avoid confusion. It must be emphasized that the model does not account for any crystalline orientation such that the boundaries between grains of the same phase are mechanically meaningless. The introduced idealization enables a clearly defined statistical analysis in which the average spatial distribution of phases around a fracture initiation site can be transparently quantified. It must be stressed that to provide a good estimate of the strain and stress fields, the numerical discretization is finer than the grid of grains, as discussed in Section~\ref{sec:numerical}. However, consistent with the adopted idealization, only grain averaged quantities are considered.

The phases are randomly distributed in each unit cell according to a predetermined volume fraction $\varphi^\mathrm{hard}$ of the hard phase. To achieve this, each individual grain in each of the $100$ unit cells is assigned the properties of the hard phase if a random number in the interval $[0,1]$ is smaller than the chosen $\varphi^\mathrm{hard}$. All other grains are assigned the properties of the soft phase. Throughout this paper, $\varphi^\mathrm{hard} = 0.25$ is used. The obtained ensemble average volume fraction of the hard phase, $\langle \varphi^\mathrm{hard} \rangle$, is within $0.1\%$ of the target value $\varphi^\mathrm{hard}$. The hard phase volume fraction in the individual unit cells varies between $\pm 3\%$ around this value.

% ==============================================================================
\subsection{Constitutive model}
\label{sec:model:cons}
% ==============================================================================

Different constitutive models provide a satisfactory description of the micromechanical response at the considered scale and level of approximation, that of an aggregate of grains. Here, both phases are modeled using small strain elasto-visco-plasticity, whereby the parameters for the viscous part are taken to effectively approximate a rate-independent model. It has been verified that this model is sufficiently accurate for the purpose of this study by comparing the response of the selected model to rate-independent models in small and finite strain; the results of this comparison are not included here for conciseness.

In the constitutive model, the linear strain tensor $\bm{\varepsilon}$ is additively split in an elastic and a plastic part:
\begin{equation}
\label{eq:model:strain_split}
  \bm{\varepsilon} =
  \bm{\varepsilon}_\mathrm{e} +
  \bm{\varepsilon}_\mathrm{p}
\end{equation}
with $\bm{\varepsilon}_\mathrm{e}$ the elastic strain tensor and $\bm{\varepsilon}_\mathrm{p}$ the plastic strain tensor.

Assuming linear elasticity, the stress tensor is determined from the elastic strain tensor by Hooke's law:
\begin{equation}
\label{eq:model:stress}
  \bm{\sigma} = \mathbb{C} : \bm{\varepsilon}_\mathrm{e}
\end{equation}
with $\bm{\sigma}$ the Cauchy stress tensor. $\mathbb{C}$ is the elastic stiffness, which is assumed to be homogeneous (independent of space) and isotropic:
\begin{equation}
  \mathbb{C} =
  K \bm{I} \otimes \bm{I} +
  2 G \big( \mathbb{I}^s - \tfrac{1}{3} \bm{I} \otimes \bm{I} \big)
\end{equation}
where the bulk modulus $K$ and the shear modulus $G$ depend on the Young's modulus $E$ and Poisson's ratio $\nu$ in the usual way: $K=E/[3(1-2\nu)]$ and $G=E/[2(1+\nu)]$. The tensors $\bm{I}$ and $\mathbb{I}^s$ respectively are the second-order identity tensor and the fourth-order symmetric identity tensor.

Plasticity is modeled using standard $J_2$ plasticity, by adopting the following yield function:
\begin{equation}
\label{eq:model:yield}
  f( \bm{\sigma} , \varepsilon_\mathrm{p} )
  =
  \sigma_\mathrm{eq} - \sigma_\mathrm{y}(\varepsilon_\mathrm{p})
\end{equation}
with $\sigma_\mathrm{eq}$ Von Mises's equivalent stress -- defined as $\sqrt{ \tfrac{3}{2} \, \bm{\sigma}_\mathrm{d} : \bm{\sigma}_\mathrm{d}}$, with $\bm{\sigma}_\mathrm{d}$ the stress deviator -- and $\sigma_\mathrm{y}$ the yield stress. The latter depends on the accumulated equivalent plastic strain
\begin{equation}
\label{eq:model:ep}
  \varepsilon_\mathrm{p}
  = \int_0^t \sqrt{
    \tfrac{2}{3} \,
    \dot{\bm{\varepsilon}}_\mathrm{p} :
    \dot{\bm{\varepsilon}}_\mathrm{p}
  } \;\mathrm{d}\tau
\end{equation}
using a linear hardening relation:
\begin{equation}
  \sigma_\mathrm{y} = \sigma_\mathrm{y0} + H \varepsilon_\mathrm{p}
\end{equation}
with $\sigma_\mathrm{y0}$ the initial yield stress and $H$ the hardening modulus.

Associative flow is assumed to model the evolution of the plastic strain rate:
\begin{equation}
\label{eq:model:plastic_rate}
  \dot{\bm{\varepsilon}}_\mathrm{p} =
  \dot{\gamma}\, \frac{\partial f}{\partial \bm{\sigma}}
\end{equation}
where the plastic multiplier is obtained through Norton's law, i.e.
\begin{equation}
\label{eq:model:norton}
  \dot{\gamma} =
  \gamma_0
  \left( \frac{\sigma_\mathrm{eq}}{\sigma_\mathrm{y}} \right)^{1/m}
\end{equation}
with $m$ the rate dependence exponent and $\gamma_0$ a material constant.

The two phases have the same elastic moduli and differ only through their plastic behavior: the hard phase has a yield stress and hardening modulus which are twice those of the soft phase. The material parameters are summarized in Table~\ref{tab:material}. Note that the rate-dependence exponent of $1/m = 100$ adequately approximates a rate-independent elasto-plastic model.

% ==============================================================================
\subsection{Applied deformation}
\label{sec:model:def}
% ==============================================================================

As a result of the assumed periodicity of the microstructure in all spatial directions, only the volume averaged deformation needs to be prescribed. In this way, local (microscopic) fluctuations in stress and strain are permitted throughout the cell and along its boundary. In the following, two different average strain paths are considered: planar and axisymmetric shear. Both strain paths are isochoric and result in a vanishing macroscopic stress triaxiality. Note that the triaxiality in the individual grains is determined by equilibrium and compatibility, and generally does not vanish.

The applied macroscopic strain tensor, $\bm{\bar{\varepsilon}}$, for planar shear reads
\begin{equation}
\label{eq:model:strain_shear}
  \bm{\bar{\varepsilon}} =
  \frac{\sqrt{3}}{2}\; \bar{\varepsilon}\,
  \left(
    \vec{e}_x \vec{e}_x -
    \vec{e}_y \vec{e}_y
  \right)
\end{equation}
in which $\bar{\varepsilon}$ is the macroscopic equivalent strain. Note that all directions are treated in the same way, fluctuations are thus allowed in the $z$-direction equivalently to the $x$- and $y$-direction. Locally the deformation is therefore three-dimensional.

For axisymmetric shear, the strain tensor reads
\begin{equation}
\label{eq:model:strain_tension}
  \bm{\bar{\varepsilon}} =
  \bar{\varepsilon}\,
  \left(
    \vec{e}_x \vec{e}_x -
    \tfrac{1}{2} \vec{e}_y \vec{e}_y -
    \tfrac{1}{2} \vec{e}_z \vec{e}_z
  \right)
\end{equation}
where $\bar{\varepsilon}$ retains its meaning of the macroscopic equivalent strain.

For both cases, the microstructure is deformed up to a value of $\bar{\varepsilon} = 0.1$. This value is large enough to be in a relevant regime for fracture initiation, but small enough to stay within the limits of the small strain approximation.

% ==============================================================================
\subsection{Fracture initiation criterion}
% ==============================================================================

The initiation of fracture at the local grain level is modeled using a damage descriptor. This descriptor reveals fracture initiation in the grains, but does not cause any material degradation. For the soft phase the Johnson-Cook model is used, which is commonly applied to study ductile fracture initiation \cite{Johnson1985}. In this model the rate of accumulated plastic strain is compared to a critical plastic strain and then integrated over the deformation history, i.e.
\begin{equation}
\label{eq:model:D}
  D =
  \int_0^t
    \frac{\dot{\varepsilon}_\mathrm{p}}{\varepsilon_\mathrm{c} (\eta) } \;
  \mathrm{d} \tau
\end{equation}
The accumulated plastic strain rate, $\dot{\varepsilon}_\mathrm{p}$, is the time derivative of \eqref{eq:model:ep} and is therefore non-negative; the critical plastic strain $\varepsilon_\mathrm{c}$ is a function of the (local) stress triaxiality $\eta$ as follows:
\begin{equation}
  \varepsilon_\mathrm{c} = A \exp ( - B \, \eta ) + \varepsilon_\mathrm{pc}
\end{equation}
The the stress triaxiality $\eta$ is the ratio of the hydrostatic stress $\sigma_\mathrm{m}$ and the equivalent stress $\sigma_\mathrm{eq}$; the material parameters $A$, $B$, and $\varepsilon_\mathrm{pc}$ are summarized in Table~\ref{tab:material}. Fracture initiation occurs in the grain once the descriptor $D \geq 1$. It is assumed that there is no fracture initiation in the hard phase.

% ------------------------------------------------------------------------------
\begin{table}[htp]
  \centering
  \caption{Material parameters of the constitutive model and the fracture initiation criterion \cite[e.g.][]{Sun2009,Al-Abbasi2003,Vajragupta2012}. Note that the phases are elastically identical, and that the hard phase is assumed not to fracture.}
  \label{tab:material}
  \vspace{0.5eM}
  \begin{tabular}{l|c|c|c|c|c|c|c|c}
          & $\nu$ & $\sigma_\mathrm{y0}/E$ & $H/E$ & $1/m$ & $\gamma_0$ & A & B & $\varepsilon_\mathrm{pc}$ \\
     \hline
     soft & 0.3 & $3 \cdot 10^{-3}$ & $8  \cdot 10^{-3}$ & 100 & 1 & 0.2 & 1.7 & 0.05 \\
     hard & 0.3 & $6 \cdot 10^{-3}$ & $16 \cdot 10^{-3}$ & 100 & 1 &  -  &  -  &  -   \\
  \end{tabular}
\end{table}
% ------------------------------------------------------------------------------

% ==============================================================================
\subsection{Fracture initiation hot-spot}
\label{sec:model:hotspot}
% ==============================================================================

The relation between the microstructure and initiation of fracture is studied by calculating the characteristic phase distribution around the fracture initiation sites. To obtain a statistically representative result, all fracture initiation sites in the entire ensemble of the $100$ random unit cells are taken into account at once. The resulting average pattern indicates what is the probability to find hard and soft phase in the vicinity of any fracture initiation site. This method, as introduced in \cite{DeGeus2015a}, is discussed in detail based on a single unit cell. The extension to the ensemble average is straightforward and therefore omitted.

A phase indicator, $\mathcal{I}$, is introduced to define the phase for each grain as a three-dimensional scalar field, i.e.\
\begin{equation}
  \mathcal{I}(i,j,k) = \begin{cases}
    1 \quad\mathrm{for}\; (i,j,k) \in \mathrm{hard} \\
    0 \quad\mathrm{for}\; (i,j,k) \in \mathrm{soft}
  \end{cases}
\end{equation}
where $(i,j,k)$ corresponds to the position in the unit cell. For the cubic microstructures considered in this paper, the index $(i,j,k)$ is the voxel index of each grain. Similarly, fracture initiation is indicated by $\mathcal{D}(i,j,k)$ which is unitary if the damage descriptor $D(i,j,k) \geq 1$, and equals zero in all other cases.

The (spatially) average(d) phase distribution around the fracture initiation sites (where $\mathcal{D} = 1$) is given by the following weighted average phase probability function:
\begin{equation}
\label{eq:hotspot}
  \mathcal{I}_\mathcal{D} (\Delta i, \Delta j, \Delta k) =
  \frac{
    \sum_{ijk} \;
    \mathcal{D}(i,j,k) \;
    \mathcal{I} ( i+\Delta i, j+\Delta j, k+\Delta k)
  }{
    \sum_{ijk} \;
    \mathcal{D} (i,j,k) \hfill
  }
\end{equation}
where $(\Delta i, \Delta j, \Delta k)$ is the relative position with respect to the initiation fracture sites, and where $(i,j,k)$ in the sums loops over all grains in the unit cell, taking the periodicity into account. Note that \eqref{eq:hotspot} effectively quantifies the cross-correlation between $\mathcal{I}$ and $\mathcal{D}$, and that it may be evaluated as a convolution (e.g.\ with FFT). It may be interpreted as the probability of finding the hard phase as a function of the position relative to the fracture initiation site. As the microstructure consists of only two phases, the probability of finding the soft phase is also contained in this result: if at a certain relative position $\mathcal{I}_\mathcal{D} ( \Delta \vec{x} ) > \varphi^\mathrm{hard}$ (the hard phase volume fraction), then hard phase at this relative position promotes fracture initiation, while if $\mathcal{I}_\mathcal{D} ( \Delta \vec{x} ) < \varphi^\mathrm{hard}$, soft phase at that relative position will promote fracture initiation.

% ==============================================================================
\section{Spectral solver for mechanical equilibrium}
\label{sec:numerical}
% ==============================================================================

% ==============================================================================
\subsection{Introduction}
% ==============================================================================

The numerical scheme introduced by Moulinec and Suquet \cite{Moulinec1998} is used in this paper. In this scheme, mechanical equilibrium is solved iteratively whereby a reference elastic medium is used to project an intermediate solution to a new, compatible, intermediate solution until equilibrium is satisfied. This method does not require the assembly of a tangent matrix or the solution of a linear system, and therefore its memory footprint scales approximately with the number of grid-points instead of its square. The projection itself is performed using the Discrete Fourier Transform which enhances the computational efficiency. Although it has been shown that the convergence rate of this particular algorithm depends critically on the choice of the reference medium, e.g.\ \cite{Brisard2010}, this issue is avoided in the present study because the microstructure is considered as elastically homogeneous. For the same reason, oscillations in the solution (e.g.\ the well-known Gibbs phenomenon) are also avoided here.

Below, we recall briefly the algorithm of Moulinec and Suquet \cite{Moulinec1998}, before explaining how the specific constitutive model enters this algorithm.

% ==============================================================================
\subsection{Solution using auxiliary problem}
% ==============================================================================

For this method, the local strains are the unknowns. The goal is therefore to find a field of periodic strains, $\bm{\varepsilon}(\vec{x})$, that satisfies mechanical equilibrium, i.e.
\begin{equation}
  \vec{\nabla} \cdot \bm{\sigma} ( \vec{x} ) = \vec{0}
\end{equation}
An auxiliary problem is introduced for which the stress $\bm{\sigma} ( \vec{x} )$ is the response of a homogeneous linear elastic body, with stiffness $\mathbb{C}^0$, subjected to a polarization stress $\bm{\tau} (\vec{x})$, i.e.
\begin{equation}
  \bm{\sigma}(\vec{x}) =
  \mathbb{C}^0 : \bm{\varepsilon}(\vec{x}) + \bm{\tau}(\vec{x})
\end{equation}
The solution to this problem is given by the following convolution, known as the periodic Lippmann-Schwinger equation:
\begin{equation}
  \bm{\varepsilon} (\vec{x}) =
  - \bm{\Gamma}^0 \star \bm{\tau}(\vec{x}) + \bar{\bm{\varepsilon}}
\end{equation}
with $\bm{\Gamma}^0$ the periodic Green operator associated with $\mathbb{C}^0$ and $\bar{\bm{\varepsilon}}$ the average strain tensor. The convolution can be evaluated in Fourier space using a simple tensor contraction, as follows:
\begin{equation}
\label{eq:num:convolution}
\begin{split}
  \hat{\bm{\varepsilon}} (\vec{q})
  &=
  - \hat{\bm{\Gamma}}^0(\vec{q}) : \hat{\bm{\tau}}(\vec{q})
  \quad\mathrm{for}\; \vec{q} \ne 0
  \\
  \hat{\bm{\varepsilon}}(\vec{0})
  &=
  \bar{\bm{\varepsilon}}
\end{split}
\end{equation}
where the $\hat{.}$ denotes that quantities have been transformed to Fourier space, and $\vec{q}$ denotes the wave-vector dependence in that space.

The polarization stress $\bm{\tau}(\vec{x})$ is generally composed of two contributions: (i) one ensuing from the inhomogeneity of elastic constants and (ii) one involving all eigenstrains such as the plastic strain $\bm{\varepsilon}_\mathrm{p} (\vec{x})$:
\begin{equation}
\label{eq:num:polarization}
  \bm{\tau}(\vec{x}) =
  \left( \mathbb{C}(\vec{x})-\mathbb{C}^0 \right) : \bm{\varepsilon} (\vec{x}) -
  \mathbb{C}(\vec{x}) : \bm{\varepsilon}_\mathrm{p} (\vec{x})
\end{equation}
Since both $\bm{\varepsilon} (\vec{x})$ and $\bm{\varepsilon}_\mathrm{p} (\vec{x})$ are not known yet, the solution has to be found iteratively. The rate of convergence is thereby determined by the choice of the reference medium (through $\bm{\Gamma}^0$).

% ==============================================================================
\subsection{Problem specific solution}
% ==============================================================================

In the present work, the microstructure is elastically homogeneous, i.e. $\mathbb{C}^0 = \mathbb{C}$ independent of $\vec{x}$, and the polarization stress becomes:
\begin{equation}
  \bm{\tau} (\vec{x})
  =
  - \mathbb{C} : \bm{\varepsilon}_\mathrm{p}^{(t+\Delta t)}
\end{equation}
where $t+ \Delta t$ denotes the current time increment. Hence, iterations can be entirely avoided by using an explicit time integration scheme for assessing the plastic strain at a new time step:
\begin{equation}
  \bm{\varepsilon}_\mathrm{p}^{(t+\Delta t)} =
  \bm{\varepsilon}_\mathrm{p}^{(t)} +
  \Delta t \; \dot{\bm{\varepsilon}}_\mathrm{p}^{(t)}
\end{equation}
substituting equations (\ref{eq:model:yield}--\ref{eq:model:norton})
\begin{align}
  \bm{\varepsilon}_\mathrm{p}^{(t+\Delta t)} &=
  \bm{\varepsilon}_\mathrm{p}^{(t)} +
  \Delta t \;
  \gamma_0 \left(
    \frac{\sigma_\mathrm{eq}^{(t)}}{\sigma_\mathrm{y}^{(t)}}
  \right)^{1/m}
  \left( \frac{\partial f}{\partial \bm{\sigma}} \right)^{(t)}
  \\
  \label{eq:num:ep-ex}
  &= \bm{\varepsilon}_\mathrm{p}^{(t)} +
  \tfrac{3}{2}  \Delta t \;
  \gamma_0 \left(
    \frac{
      \sigma_\mathrm{eq}^{(t)}
    }{
      \sigma_\mathrm{y0} + H \varepsilon_\mathrm{p}^{(t)}
    }
  \right)^{1/m}
  \frac{\bm{\sigma}_\mathrm{d}^{(t)}}{\sigma_\mathrm{eq}^{(t)}}
\end{align}

Finally, knowing the polarization stress, its Fourier transform is inserted into \eqref{eq:num:convolution} to determine the Fourier transform of the strain field at the new time increment, $\hat{\bm{\varepsilon}}^{(t+\Delta t)}$. The inverse Fourier transform then provides the corresponding strain field in real space, $\bm{\varepsilon}^{(t+\Delta t)}$, which were the unknowns.

% ==============================================================================
\subsection{Applied discretization}
% ==============================================================================

The Fourier transform and its inverse are evaluated using the Discrete Fourier Transform, for which efficient open source implementations are readily available. Each of the grains of the microstructure is discretized using $5 \times 5 \times 5$ grid points (see Figure~\ref{fig:microstructure}(a)). It has been verified that this discretization leads to a converged solution in terms of the grain averaged tensors used in this paper. The refinement study confirmed that the approximation is smaller than 1\% relative error in stress and (plastic) strain measured in the individual grains.

To minimize the approximation of the explicit time integration, the deformation is applied in $10^5$ small steps. It has been verified that this temporal discretization is small enough to ensure stability and convergence of the scheme. Note that even though many steps are applied, the computations at each time step are fast, even for the large three-dimensional unit cells.

% ==============================================================================
\section{Two-dimensional vs.\ three-dimensional microstructure}
\label{sec:res:microstructure}
% ==============================================================================

The effect of the three-dimensional (3-D) nature of the considered microstructures is investigated by comparing them to their two-dimensional (2-D) counterpart for which each cross-section in $z$-direction is the same. In this comparison both the 2-D and 3-D microstructures are subjected to macroscopic planar shear (see Equation~\ref{eq:model:strain_shear}). Since the number of considered morphologies is smaller for the 2-D microstructures, the size of the ensemble is increased to $256$ random unit cells.

% ==============================================================================
\subsection{Macroscopic response}
% ==============================================================================

\begin{figure}[tph]
  \centering
  \includegraphics[width=0.6\linewidth]{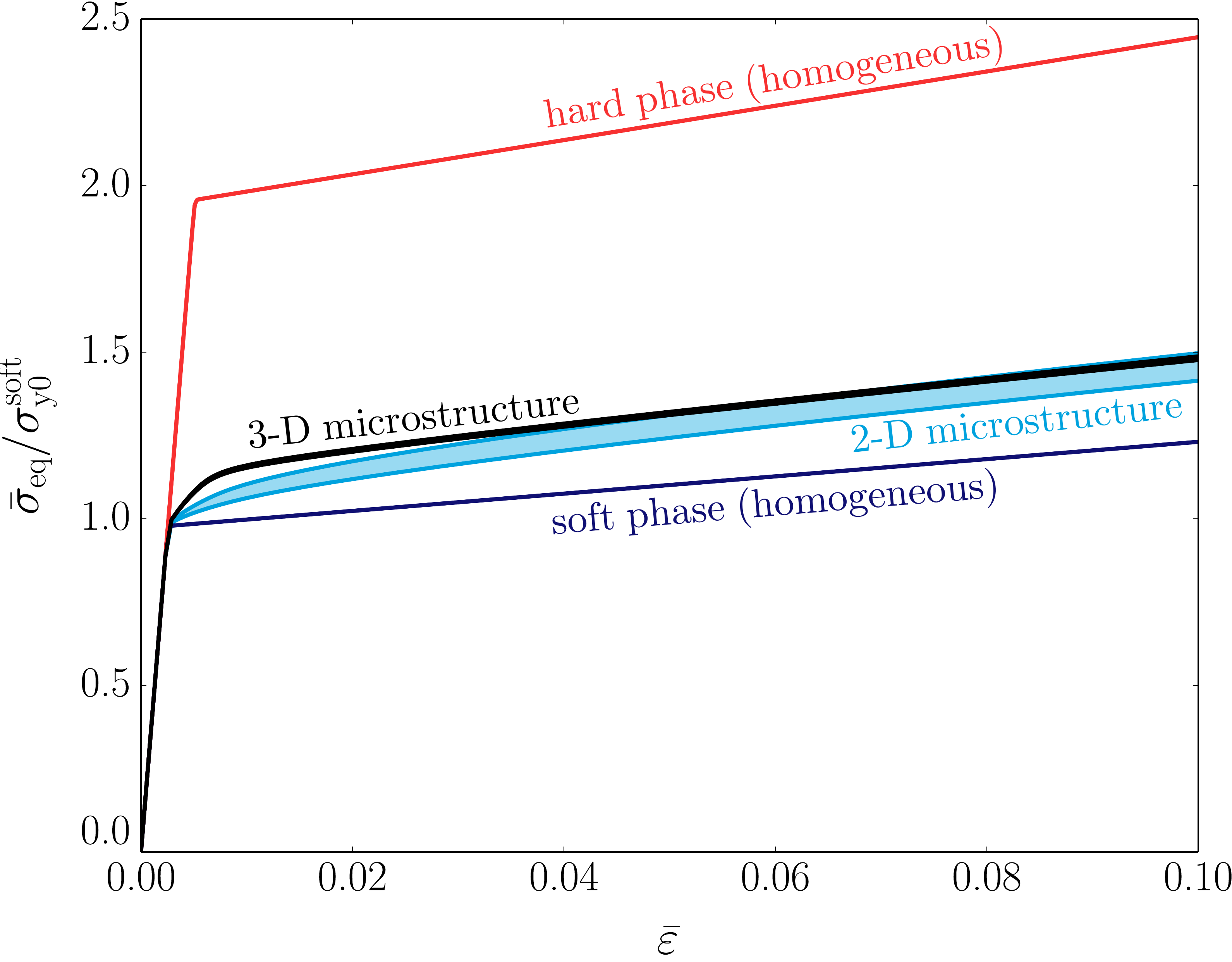}
  \caption{The macroscopic equivalent stress, $\bar{\sigma}_\mathrm{eq}$, as a function of the applied equivalent strain, $\bar{\varepsilon}$. The range of the responses of the unit cells is shown for the 2-D microstructures (blue) and the 3-D microstructures (black), both for applied planar shear (Eq.~\ref{eq:model:strain_shear}). The constitutive responses of the hard phase (red) and the soft phase (dark blue) are shown as reference. }
  \label{fig:macroscopic_pureshear}
\end{figure}

The computed macroscopic equivalent stress response, $\bar{\sigma}_\mathrm{eq}$, as a function of the applied equivalent strain, $\bar{\varepsilon}$, is shown in Figure~\ref{fig:macroscopic_pureshear}. For the 3-D microstructures, in black, and the 2-D microstructures, in blue, the upper and lower bounds of the different unit cells are plotted. The constitutive response of the individual phases is also shown using a dark blue curve for the soft matrix and red for the hard reinforcement phase. As observed, all curves coincide in the elastic regime, as the microstructure is elastically homogeneous. In the plastic regime, the response of the microstructure is a non-linear combination of the response of the individual phases. The spread between the responses of the different unit cells is larger for the 2-D model than for the 3-D model. The fact that the scatter in volume fraction between the 2-D unit-cells is somewhat larger (it is in the range $0.21 < \varphi^\mathrm{hard} < 0.29$) thereby only partly explains the difference. A clear difference is also observed in the plastic response directly after yielding. The response of the 2-D model is well below that of the 3-D model. The difference is explained by examining the local plastic response in the next section.

% ------------------------------------------------------------------------------
% ------------------------------------------------------------------------------

% ==============================================================================
\subsection{Microscopic response}
% ==============================================================================

\begin{figure}[htp]
  \centering
  \includegraphics[width=0.7\linewidth]{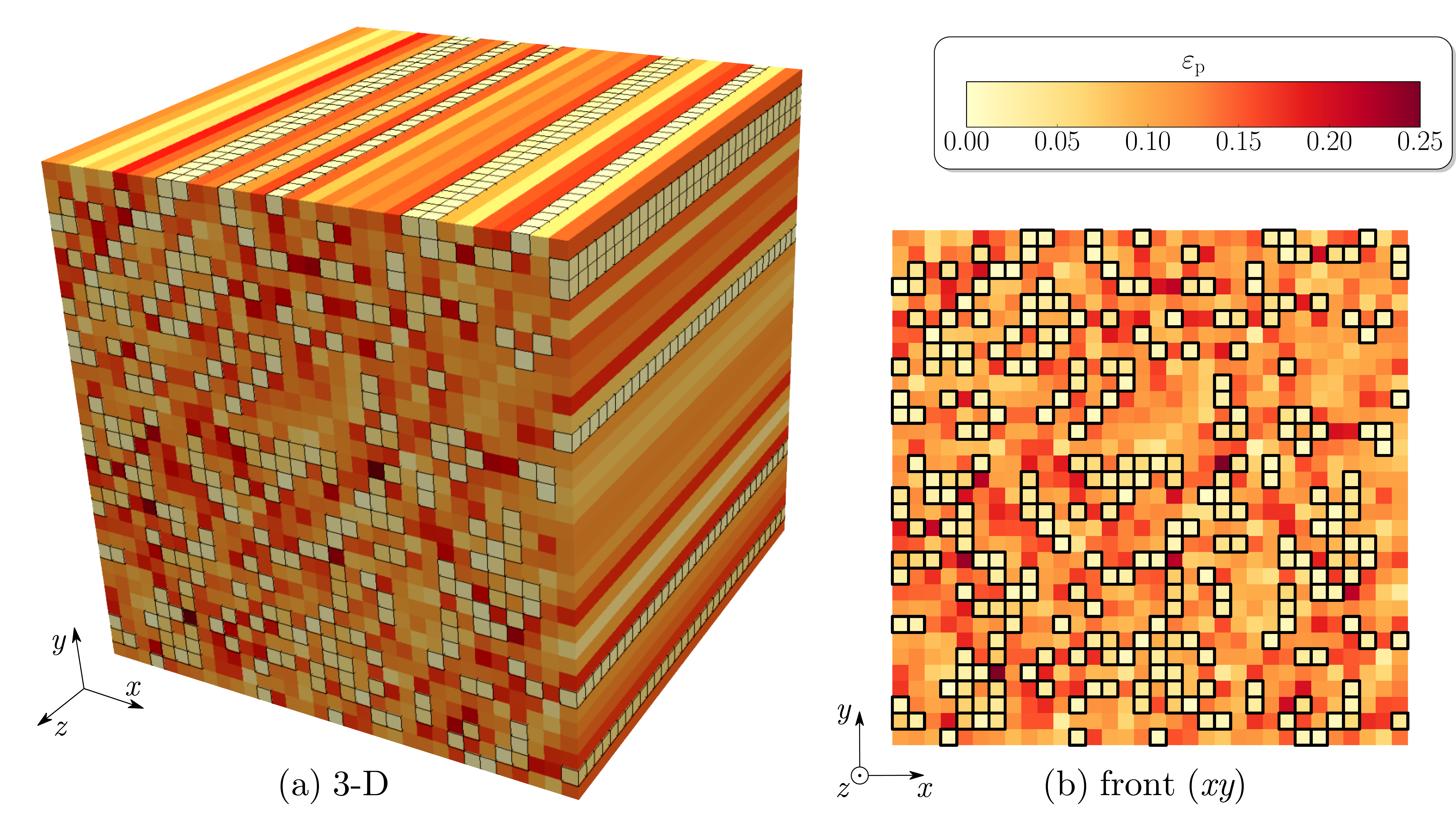}
  \caption{Local accumulated plastic strain, $\varepsilon_\mathrm{p}$, in the 2-D microstructure, resulting from an applied macroscopic planar shear strain with equivalent value $\bar{\varepsilon} = 0.1$. The hard grains are indicated using a black outline. The different views: (a) a three-dimensional view, and (b) the cross-section along the $xy$-plane.}
  \label{fig:ep_typical_2D}
\end{figure}

\begin{figure}[htp]
  \centering
  \includegraphics[width=0.75\linewidth]{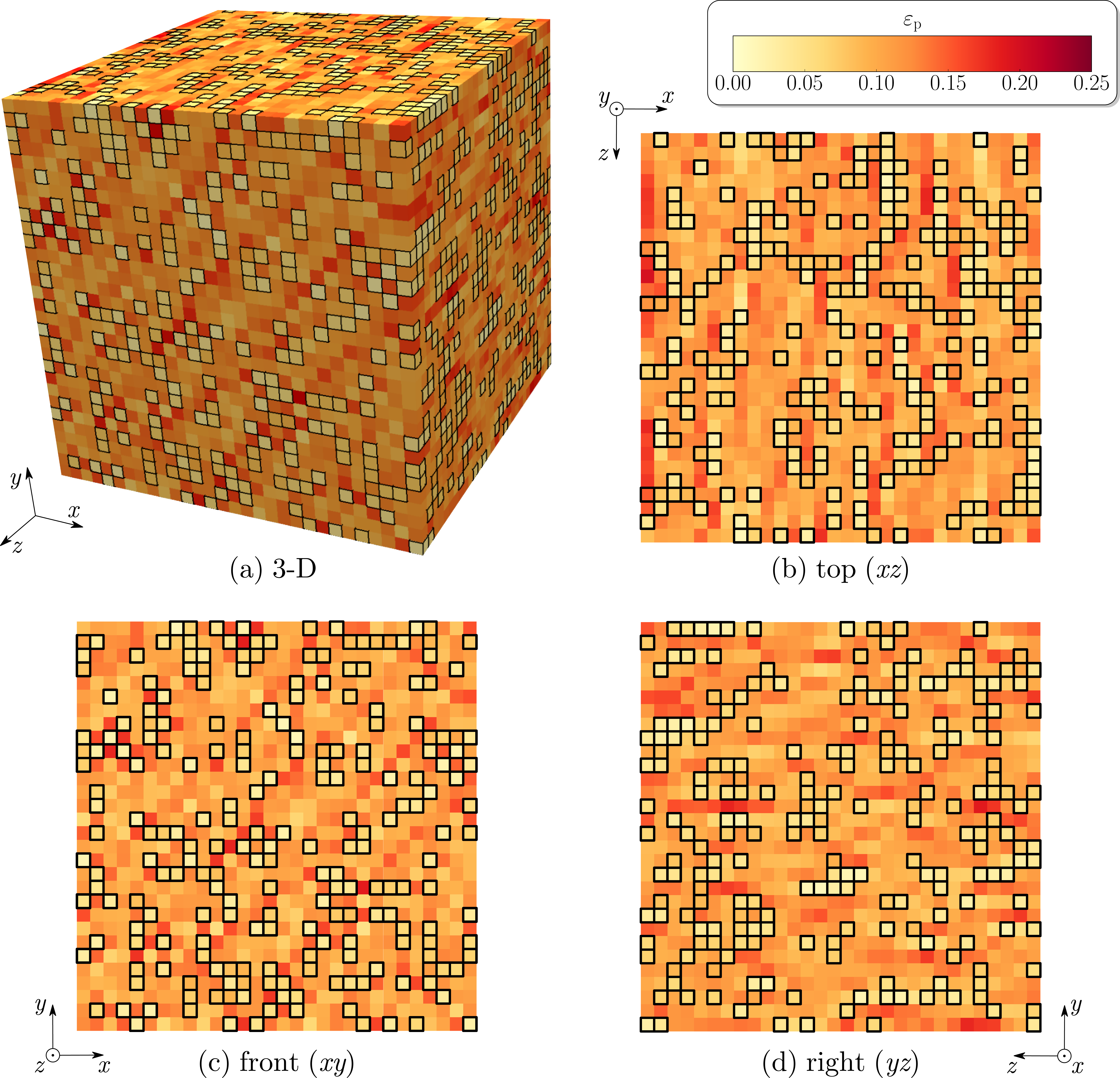}
  \caption{Local accumulated plastic strain, $\varepsilon_\mathrm{p}$, in the 3-D microstructure (of Figure~\ref{fig:microstructure}); cf.\ Figure~\ref{fig:ep_typical_2D}. The different views: (a) a three-dimensional view, (b--d) each of the visible cross-sections.}
  \label{fig:ep_typical_shear}
\end{figure}

The local accumulated plastic strain, $\varepsilon_\mathrm{p}$, is shown in Figure~\ref{fig:ep_typical_2D} for the 2-D microstructure and in Figure~\ref{fig:ep_typical_shear} for the 3-D microstructure; in both cases for an applied equivalent strain $\bar{\varepsilon} = 0.1$. The figures include a three-dimensional view of the unit cell and planar views of the visible (outer) planes (for the 2-D microstructure only one view is relevant). Note that because of the periodicity of the microstructure, the outer planes are in fact cross-sections of the material.

For both 2-D and 3-D microstructures, it is observed that the highest values of $\varepsilon_\mathrm{p}$ occur in the soft phase (with an average of $0.13$ and $0.12$ respectively). In most of the hard grains $\varepsilon_\mathrm{p}$ is considerably lower (with an average of $0.02$ and $0.04$ respectively). From the cross-section along the $xy$-direction (Figures~\ref{fig:ep_typical_2D}(b) and \ref{fig:ep_typical_shear}(c)), it is observed that the highest values of $\varepsilon_\mathrm{p}$ are localized in bands of connected soft phase at $\pm 45$ degree angles with respect to the $x$-axis, in particular if these bands are surrounded by hard phase. For the 3-D microstructure, the cross-sections along the $xz$-plane and the $yz$-plane (Figure~\ref{fig:ep_typical_shear}(b,d)) reveal that these shear bands are extended along the $z$-axis.

Plastic strain thus localizes in shear bands at $\pm 45$ degree angles with respect to the $x$- and $y$-axis. This coincides with the direction of maximum shear of the applied macroscopic strain. It can be concluded that the highest plastic strain occurs where the soft phase is connected in the direction of shear, and surrounded by the hard phase.

\begin{figure}[htp]
  \centering
  \includegraphics[width=0.8\linewidth]{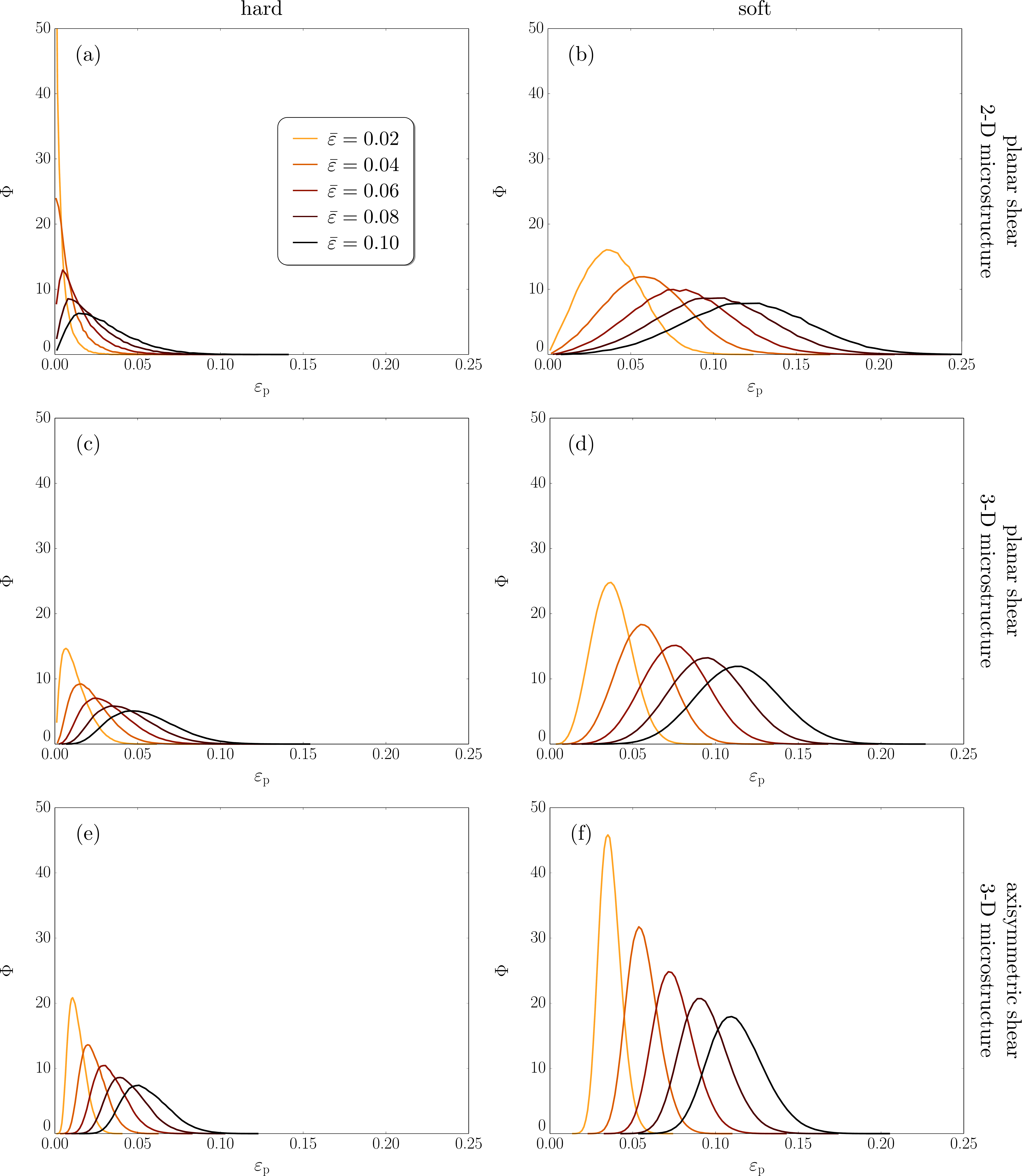}
  \caption{Probability density, $\Phi$, of the accumulated plastic strain, $\varepsilon_\mathrm{p}$, at different strain increments. The rows correspond to: planar shear on a 2-D microstructure (a--b), planar shear on a 3-D microstructure (c--d), and axisymmetric shear on a 3-D microstructure (e--f) at different strain increments. The columns refer to hard phase (left) and to soft phase (right).}
  \label{fig:ep_pdf_inc_3D}
\end{figure}

These observations, made on a single unit cell, can be extended to the entire ensemble. In Figure~\ref{fig:ep_pdf_inc_3D} the probability density, $\Phi$, of the accumulated plastic strain, $\varepsilon_\mathrm{p}$, is shown for different strain increments. The figure includes both the 2-D (Figure~\ref{fig:ep_pdf_inc_3D}(a--b)) and 3-D microstructures (Figure~\ref{fig:ep_pdf_inc_3D}(c--d)); the hard phase (left column) and the soft phase (right column) are shown separately. The value of $\Phi$, along the vertical axes, is normalized such that the area underneath the curve is unity. The results indicate that, throughout the deformation history, the strain is more strongly partitioned between the soft and the hard phase for the 2-D microstructure than for the 3-D microstructure (i.e.\ larger plastic strains, mainly in the soft phase). This explains why the macroscopic hardening mostly originates from the soft phase for the 2-D microstructure. Furthermore, the plastic strain distribution is more heterogeneous for the 2-D microstructure, because any local phase arrangement that promotes plastic straining extends infinitely in the out-of-plane direction (i.e.\ bands of soft phase at $\pm 45^\circ$). In the 3-D case, the sub-surface microstructure restricts the deformation, resulting in a less heterogeneous distribution of plastic strain.

\begin{figure}[htp]
  \centering
  \includegraphics[width=1.0\linewidth]{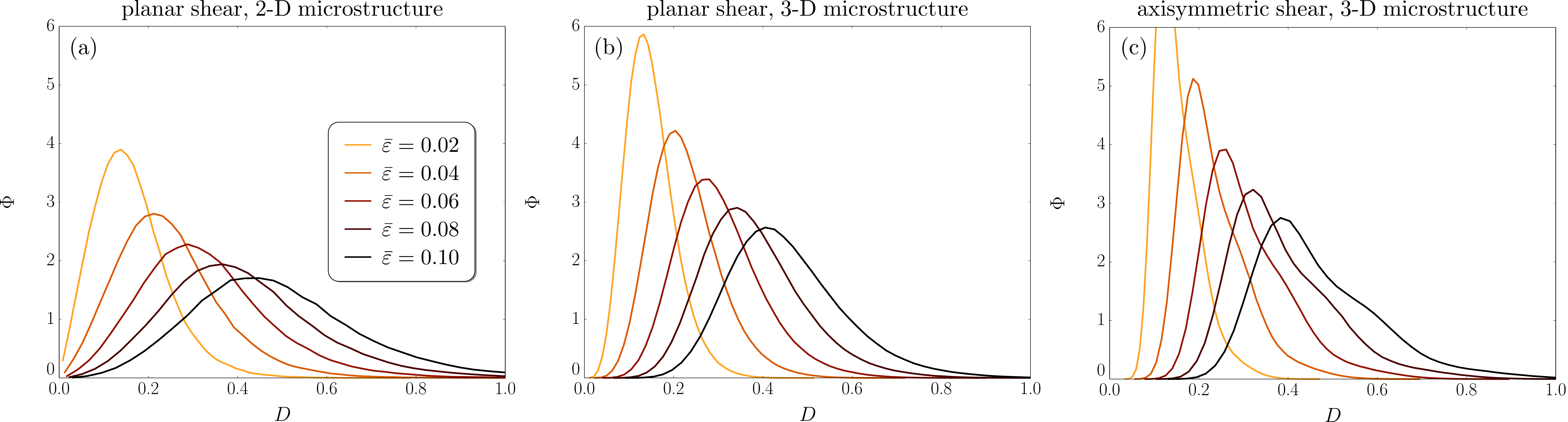}
  \caption{Probability density, $\Phi$, of the ductile damage descriptor, $D$, in the soft phase at different strain increments; for (a--b) planar shear on a two- and three-dimensional microstructure respectively, and (c) axisymmetric shear on a three-dimensional microstructure.}
  \label{fig:D_pdf_inc_3D}
\end{figure}

A similar observation holds for the damage descriptor in the soft phase, shown in Figure~\ref{fig:D_pdf_inc_3D}(a--b). For the 2-D microstructures, more grains have a higher value of $D$. More importantly, also more grains display a value of $D \geq 1$, indicating fracture initiation. In other words, fracture initiation is `promoted' in a two-dimensional microstructure, for the same reasons as observed above.

% ==============================================================================
\subsection{Hot-spot for fracture initiation}
% ==============================================================================

\begin{figure}[tph]
  \centering
  \includegraphics[width=0.45\linewidth]{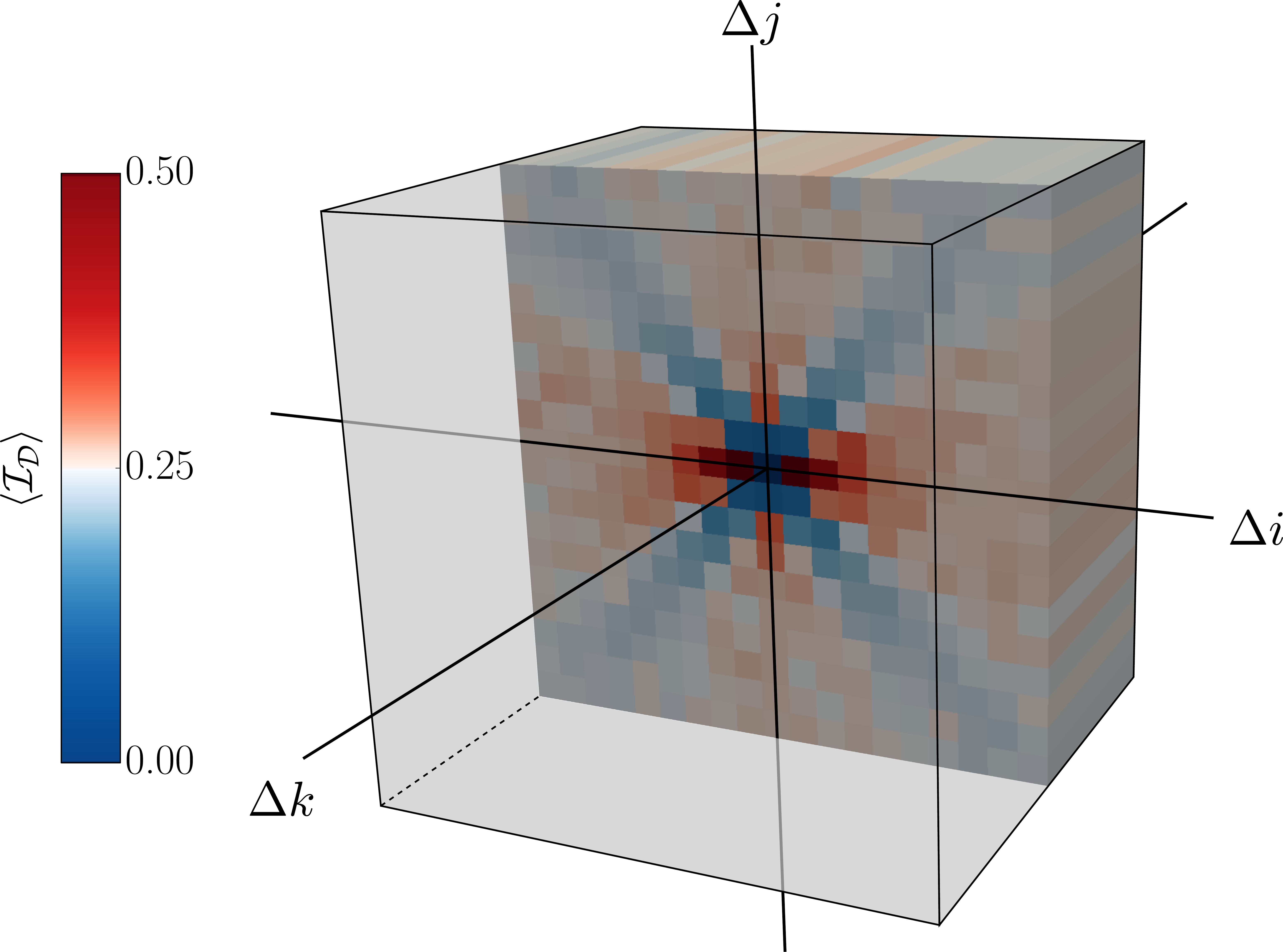}
  \caption{Fracture initiation hot-spot resulting from the 2-D microstructures loaded in planar shear. The color indicates the probability of finding the hard phase, $\langle \mathcal{I}_\mathcal{D} \rangle$, at a position $(\Delta i, \Delta j, \Delta k)$ relative to fracture initiation.}
  \label{fig:hotspot_pureshear_2D}
\end{figure}

\begin{figure}[tph]
  \centering
  \includegraphics[width=1.0\linewidth]{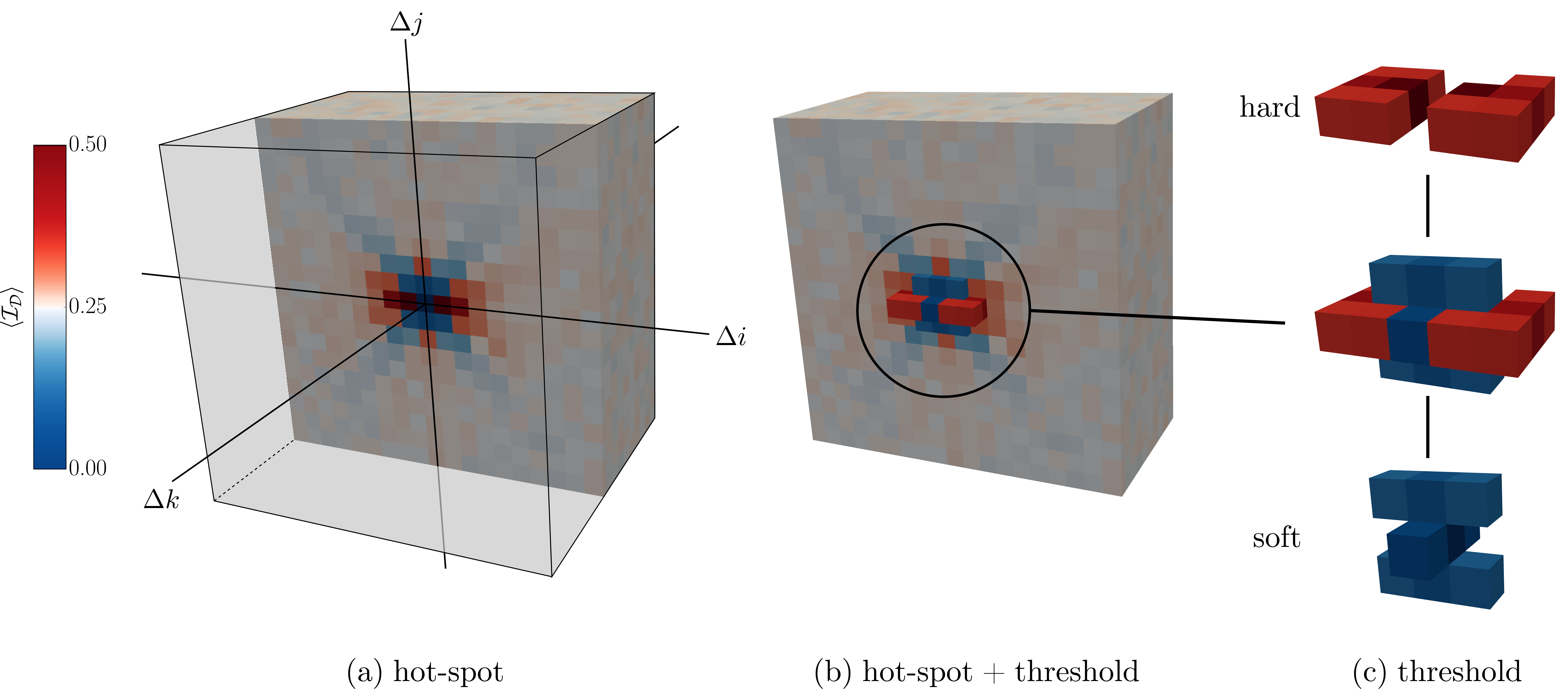}
  \caption{Fracture initiation hot-spot resulting from the 3-D microstructures loaded in planar shear; cf.\ Figure~\ref{fig:hotspot_pureshear_2D}. From left to right: (a) a cross-section parallel to the $xy$-plane, (b) the same cross-section with an applied threshold $\big| \langle \mathcal{I}_\mathcal{D} \rangle - \varphi^\mathrm{hard} \big| \geq 0.05$, and (c) the isolated threshold result.}
  \label{fig:hotspot_pureshear_3D}
\end{figure}

The fracture initiation hot-spot is statistically identified by considering all unit cells. As discussed in Section~\ref{sec:model:hotspot}, this leads to the probability of finding the hard phase at a certain position relative to fracture initiation: $\langle \mathcal{I}_D \rangle (\Delta i, \Delta j, \Delta k)$. The relative voxel coordinates $(\Delta i, \Delta j, \Delta k)$ are thereby aligned with the spatial coordinates $(x,y,z)$. The result is shown in Figure~\ref{fig:hotspot_pureshear_2D} for the 2-D microstructure and in Figure~\ref{fig:hotspot_pureshear_3D} for the 3-D microstructure. Both of these figures include a cross-section along the $xy$-plane. For the 3-D microstructure a threshold $\big| \langle \mathcal{I}_\mathcal{D} \rangle - \varphi^\mathrm{hard} \big| \geq 0.05$ is applied in order to visualize the most important morphological patterns. The probability scale-bar is taken symmetric about the hard phase volume fraction $\varphi^\mathrm{hard} = 0.25$ to avoid an optical bias. The results may be interpreted as follows: $\varphi^\mathrm{hard} < \langle \mathcal{I}_D \rangle \leq 1$ (red) indicates an elevated probability of finding the hard phase at a position relative to the fracture initiation sites and $0 \leq \langle \mathcal{I}_D \rangle < \varphi^\mathrm{hard}$ (blue) an elevated probability of finding the soft phase there.

For the 2-D microstructure, Figure~\ref{fig:hotspot_pureshear_2D} shows that, as assumed at the onset, fracture initiates in the soft phase which is indicated by the blue grain at $(\Delta i, \Delta j, \Delta k) = (0,0,0)$. Directly next to fracture initiation in $x$-direction hard phase is found in most cases, since $\langle \mathcal{I}_\mathcal{D} \rangle = 0.8$. Perpendicular, in $y$-direction, the soft phase is found with $\langle \mathcal{I}_\mathcal{D} \rangle = 0.1$. Further away, in $x$-direction regions of the hard phase are found on both sides. The regions of the hard phase are interrupted by regions of the soft phase under $\pm 45$ degree angles with respect to the $x$-axis. This result is consistent with \cite{DeGeus2015a} even though a different constitutive model and damage indicator were used there.

For the 3-D microstructure, the cross-section along the $xy$-plane in Figure~\ref{fig:hotspot_pureshear_3D}(a) closely resembles the result of the 2-D microstructure. The probability is higher of finding hard phase ($\langle\mathcal{I_D}\rangle=0.99$) left and right, and soft phase ($\langle\mathcal{I_D}\rangle=0.08$) above and below. Figure~\ref{fig:hotspot_pureshear_3D}(c) shows that fracture initiates where a `plate' of hard phase is interrupted by two `channels' of soft phase. Soft phase under $\pm 45$ degree angles is only recovered along the in-plane direction.

% ------------------------------------------------------------------------------
% ------------------------------------------------------------------------------

% ------------------------------------------------------------------------------
% ------------------------------------------------------------------------------

% ==============================================================================
\section{Two-dimensional vs.\ three-dimensional deformation}
\label{sec:res:deformation}
% ==============================================================================

The two different deformation paths from Section~\ref{sec:model:def} are compared next, both applied to 3-D microstructures.

% ==============================================================================
\subsection{Macroscopic response}
% ==============================================================================

\begin{figure}[tph]
  \centering
  \includegraphics[width=0.6\linewidth]{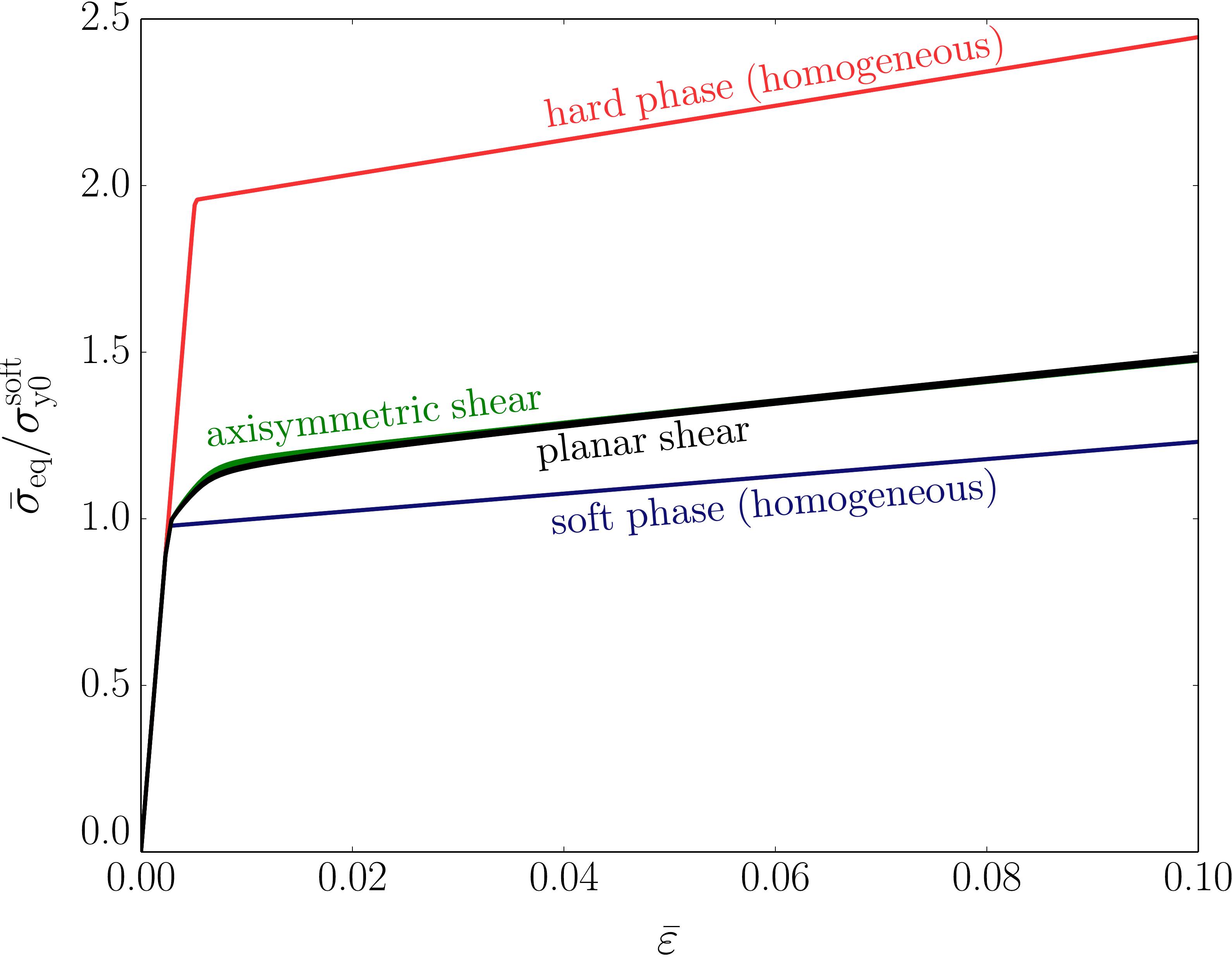}
  \caption{The macroscopic equivalent stress response, $\bar{\sigma}_\mathrm{eq}$, normalized by the initial yield stress of the soft phase, $\sigma_\mathrm{y0}^\mathrm{soft}$, as a function of the applied equivalent strain $\bar{\varepsilon}$ of all microstructures, for: planar shear (black) and axisymmetric shear (green). The bounds representing the soft (blue) and hard phase (red) are also depicted.}
  \label{fig:macroscopic_3D}
\end{figure}

The macroscopic responses are shown in Figure~\ref{fig:macroscopic_3D}. The  black curves correspond to applied planar shear and the green to applied axisymmetric shear, in between the responses of the individual phases (blue for soft and red for hard). There is little scatter, below 1\%, for the different microstructures in the ensemble, i.e.\ the upper and lower bounds of all unit cells almost coincide. At the onset of yielding, slightly more hardening is observed under axisymmetric shear than for planar shear, as the green curves lie just above the black curves for $\bar{\varepsilon} < 0.05$ (the difference is so small that it is difficult to see).

% ==============================================================================
\subsection{Microscopic plastic response}
% ==============================================================================

\begin{figure}[htp]
  \centering
  \includegraphics[width=0.75\linewidth]{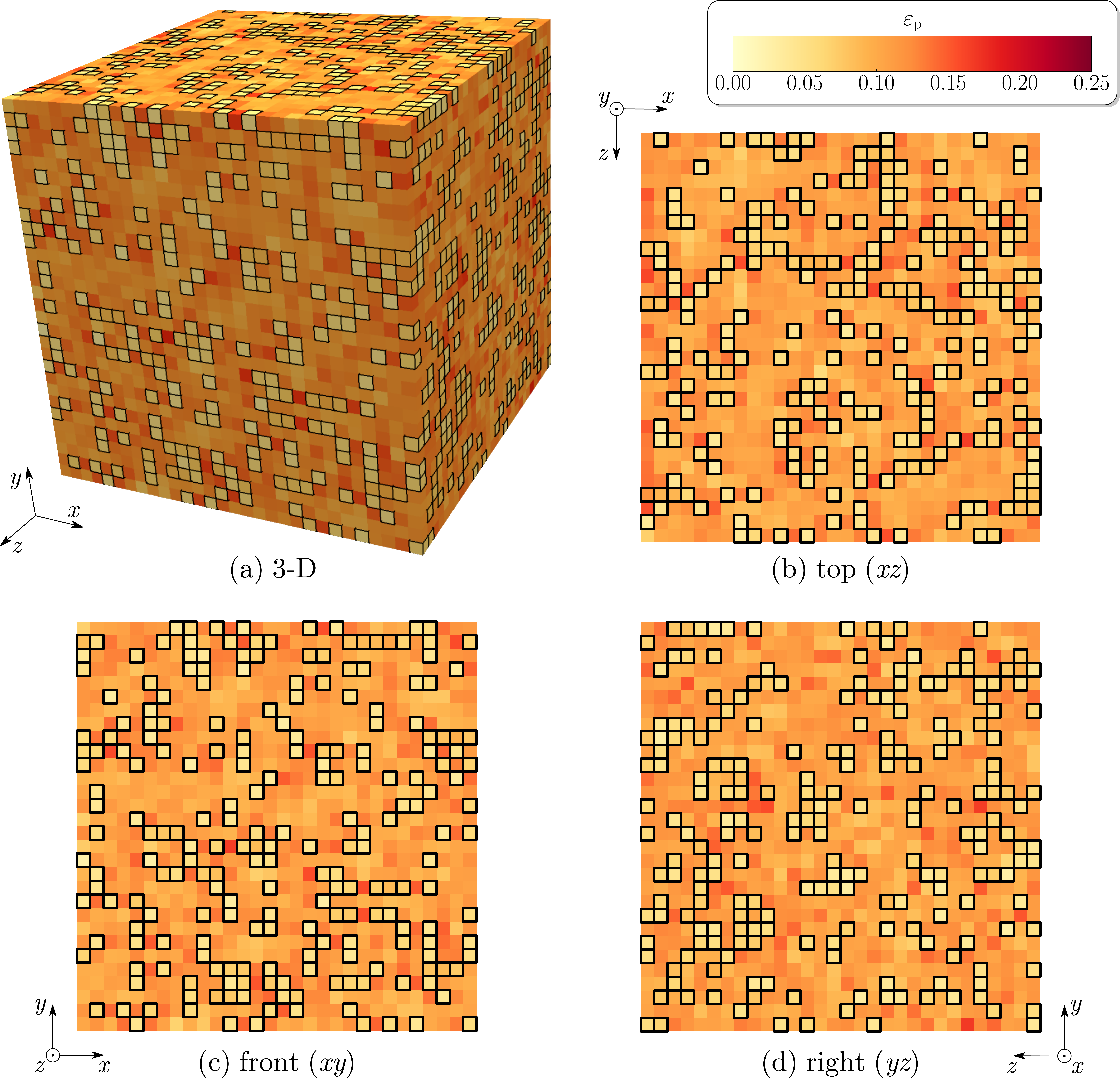}
  \caption{Local accumulated plastic strain, $\varepsilon_\mathrm{p}$, resulting from an applied macroscopic axisymmetric shear strain with equivalent value $\bar{\varepsilon} = 0.1$. The microstructure and cross-sections are identical to Figure~\ref{fig:ep_typical_shear}.}
  \label{fig:ep_typical_tension}
\end{figure}

In addition to the response to planar shear, in Figure~\ref{fig:ep_typical_shear}, the response to axisymmetric shear is shown in Figure~\ref{fig:ep_typical_tension}. It is observed that the plastic strain response is more homogeneous for this case. Less extreme values are observed in the soft phase, and also the average of $0.11$ is smaller than for planar shear. At the same time the plastic strain is higher in the hard phase, where the average value is $0.05$. In each of the three cross-sections, small regions of elevated $\varepsilon_\mathrm{p}$ occur in the soft phase, at $\pm 45$ degree angles with respect to the three axes. These localized regions are located where soft phase is surrounded by hard phase. In addition, isolated grains are observed in the soft phase where $\varepsilon_\mathrm{p}$ is high. These regions are part of a previously described shear band along the in-plane direction.

The probability density, $\Phi$, of the accumulated plastic strain, $\varepsilon_\mathrm{p}$ is shown in Figures~\ref{fig:ep_pdf_inc_3D}(e--f) for axisymmetric shear. In comparing to the earlier results for planar shear (cf.\ to Figures~\ref{fig:ep_pdf_inc_3D}(c--d)) it is observed that, throughout the deformation history, the plastic strain is more heterogeneous for planar shear than for axisymmetric shear. For the soft phase, Figure~\ref{fig:ep_pdf_inc_3D}(d,f), a larger number of grains display a high plastic strain, and also the maximum is higher. For the hard phase, Figure~\ref{fig:ep_pdf_inc_3D}(c,e), it is observed that the average plastic strain is higher for axisymmetric shear than for planar shear. Furthermore, the distribution is more skewed towards low values of $\varepsilon_\mathrm{p}$ for planar shear, indicating that most of the hard grains reveal a low $\varepsilon_\mathrm{p}$.

% ==============================================================================
\subsection{Ductile fracture initiation}
% ==============================================================================

\begin{figure}[tph]
  \centering
  \includegraphics[width=1.0\linewidth]{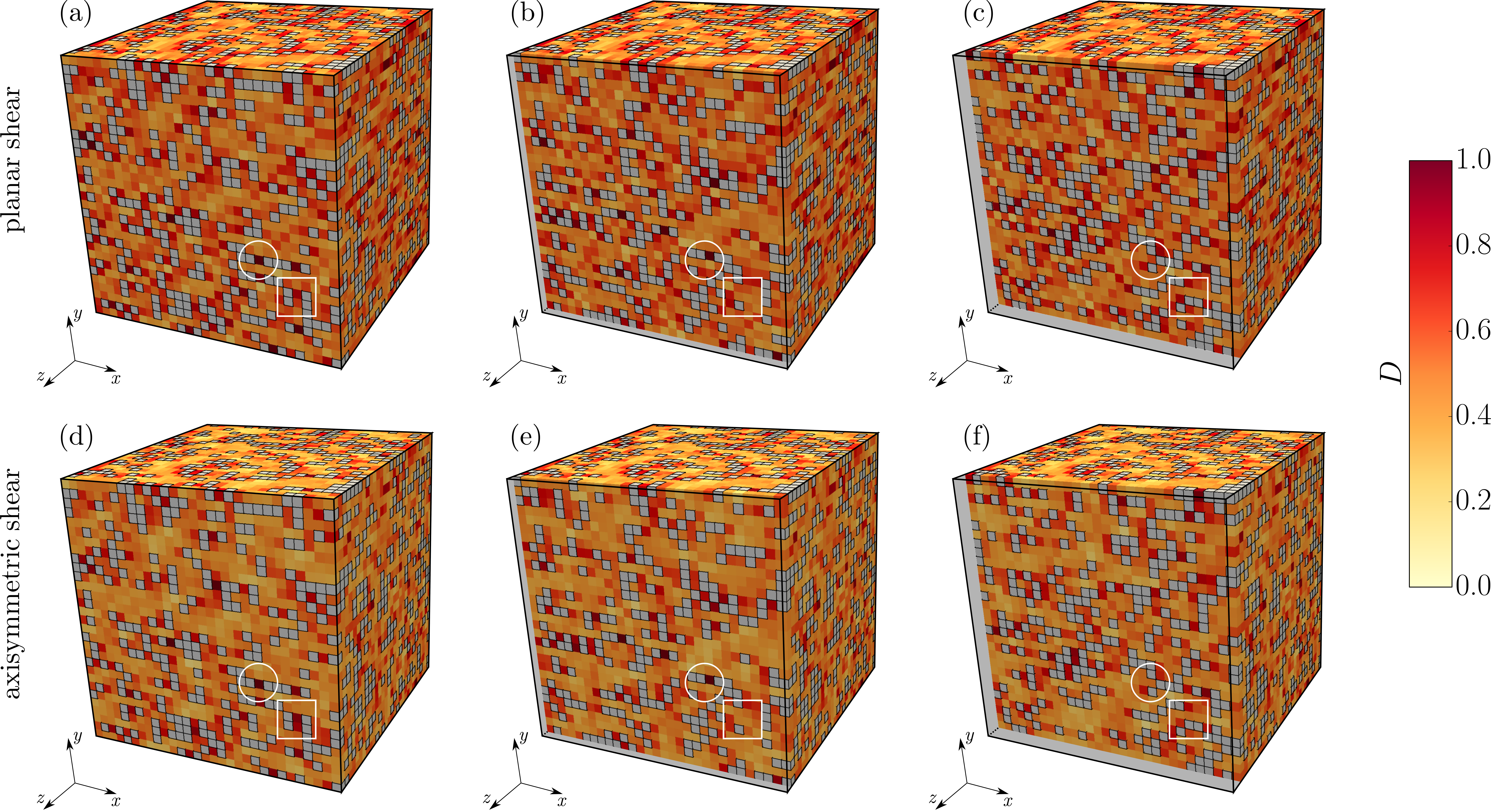}
  \caption{Damage descriptor for an applied equivalent strain of $\bar{\varepsilon} = 0.1$ for planar shear (a--c) and axisymmetric shear (d--f). From left to right the depth of the cross-section increases by one grain. The hard phase grains have a black outline, the gray value in these grains denote that fracture initiation of the hard phase is not considered.}
  \label{fig:damage_typical}
\end{figure}

Figure~\ref{fig:damage_typical} shows the damage descriptor, $D$, in the microstructure of Figure~\ref{fig:microstructure} for an applied equivalent strain of $\bar{\varepsilon} = 0.1$ for both load cases. The top row, Figure~\ref{fig:damage_typical}(a--c), shows three different cross-sections parallel to the $xy$-plane for the applied planar shear case. The bottom row, Figure~\ref{fig:damage_typical}(d--f), shows the same cross-sections for the axisymmetric shear case. The different cross-sections are taken at respectively one and two grains in negative $z$-direction with respect to the reference view on the left. The hard grains are colored gray, as damage in the hard phase is not considered.

For axisymmetric shear it is observed that the damage, $D$, is lower in many of the grains when compared with the planar shear case (cf.\ Figures~\ref{fig:damage_typical}(a,d) for a single unit cell and Figures~\ref{fig:D_pdf_inc_3D}(b,c) for the ensemble). This is a direct consequence of the less heterogeneous distribution of the plastic strain. The purpose of the damage descriptor is however to identify fracture initiation, i.e.\ the focus should be on the sites where $D \geq 1$. The number of these fracture initiation sites is approximately the same (cf.\ Figures~\ref{fig:D_pdf_inc_3D}(b,c)). A large part of the fracture initiation sites coincide for the two load cases (e.g.\ the grain highlighted with a white circle), but also clear differences exist. One example is highlighted with a white square, for which the damage is high for axisymmetric shear, but not for planar shear (cf.\ Figures~\ref{fig:damage_typical}(a,d)). For this case, the morphology consisting of a soft grain flanked by hard grains on both sides in $x$-direction and soft grains on both sides in $y$-direction is not extended into the subsurface (Figure~\ref{fig:damage_typical}(e)). This shows that spatial phase distribution, directly adjacent to but also further away from the fracture initiation site, in combination with the deformation, has to be critical for fracture to initiate.

% ==============================================================================
\subsection{Hot-spot for fracture initiation}
% ==============================================================================

\begin{figure}[tph]
  \centering
  \includegraphics[width=1.0\linewidth]{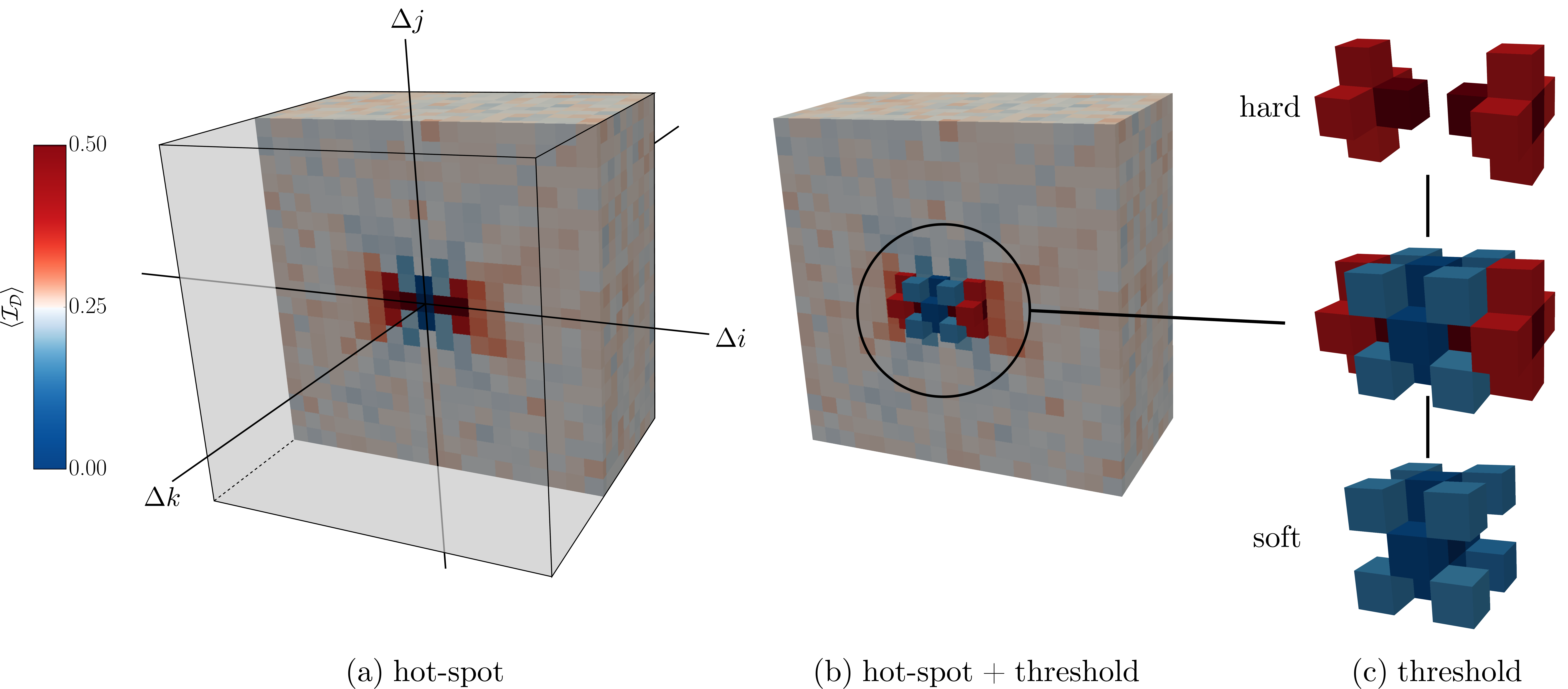}
  \caption{The fracture initiation hot-spot resulting form applied planar shear. The views are identical to those used in Figure~\ref{fig:hotspot_pureshear_3D}.}
  \label{fig:hotspot_uniaxial_3D}
\end{figure}

The fracture initiation hot-spot for axisymmetric shear is shown in Figure~\ref{fig:hotspot_uniaxial_3D}, with the same views as used in Figure~\ref{fig:hotspot_pureshear_3D}. In the case of axisymmetric shear, the cross-section in Figure~\ref{fig:hotspot_uniaxial_3D}(a) reveals a pattern that is qualitatively the same as above. Adjacent to the soft grain in the center, in $x$-direction, hard phase is identified ($\langle \mathcal{I}_\mathcal{D} \rangle = 1.0$), and in $y$-direction soft phase is found in most of the cases ($\langle \mathcal{I}_\mathcal{D} \rangle = 0.03$). The influence of the morphology next to the fracture initiation site is thereby more pronounced than for planar shear. Further away, regions of hard phase are found in $x$-direction, which are interrupted by regions of the soft phase at $\pm 45$ degrees with respect to the $x$-axis. The latter are however less prominent than for planar shear. Extending the observations to three dimensions in Figure~\ref{fig:hotspot_uniaxial_3D}(b--c), the regions of hard phase in $x$-direction tend to be axisymmetric around the $x$-axis. Regions of the soft phase are found in cones at $\pm 45^\circ$ around the $x$-axis. Note that for all neighboring voxels at $\pm 45^\circ$ the resulting $\langle \mathcal{I}_\mathcal{D} \rangle$ indicates soft phase, but for some it is below the threshold value.

In earlier work, De Geus et al.\ \cite{DeGeus2015a} used a simple mechanical analysis to confirm that the combination of a band of hard phase in the direction of applied tension, in combination with a phase boundary perpendicular to it, results in a hydrostatic tensile stress state. The regions of the soft phase in the directions of principal shear induce a large plastic strain. The combination of these morphological features promotes fracture initiation. This also explains the results for the three-dimensional microstructure and deformation. For planar shear, in Figure~\ref{fig:hotspot_pureshear_3D}, the plate of hard phase corresponds to a phase boundary whose area perpendicular to the tensile axis is maximized in such a way that the soft regions at $\pm 45$ degrees with respect to the $x$-axis are uninterrupted. The principal shear acts only in the $xy$-plane and therefore soft phase is only recovered in that plane. For axisymmetric shear, the principal shear acts on all planes at $\pm 45$ degree angles around fracture initiation, explaining the resulting morphology of soft phase. The observed pattern of hard phase again maximizes the area of the phase boundary perpendicular to the axis of applied tension avoiding interruption of the shear bands.

% ==============================================================================
\section{Concluding remarks}
\label{sec:conclusion}
% ==============================================================================

A two-phase microstructure is studied for which the two phases have different mechanical properties (yield stresses), resulting in a sharp mechanical contrast. Consequently, fracture initiation in the soft phase is strongly influenced by the spatial distribution of the phases. A large ensemble of random three-dimensional microstructures is analyzed to systematically characterize the configuration that is most critical for fracture initiation. To this end, all fracture initiation sites are considered at once by calculating the average phase distribution around fracture initiation sites.

Fracture preferentially initiates when the following conditions apply:
\begin{enumerate}
  \item regions of soft phase are aligned with the shear directions;
  \item phase boundaries perpendicular to the direction of principal strain, i.e.\ hard phase on both sides, aligned with the tensile axis;
  \item regions of hard phase are located next to the shear bands.
\end{enumerate}

Using a two-dimensional microstructure, the same conclusions are obtained. Indeed, the average phase distribution is a cross-section of the result obtained in 3-D. However, the strain is more strongly partitioned between the phases as the deformation is not restricted or promoted by the sub-surface microstructure. As a result, fracture initiation is more frequently observed in 2-D at the same overall strain.

% ==============================================================================
\section*{Acknowledgments}
% ==============================================================================

This research was carried out under project number M22.2.11424 in the framework of the research program of the Materials innovation institute M2i (\href{http://www.m2i.nl}{www.m2i.nl}).

% ==============================================================================
\section*{References}
% ==============================================================================

% ==============================================================================
\scriptsize
\bibliography{library}
% ==============================================================================

% %%%%%%%%%%%%%%%%%%%%%%%%%%%%%%%%%%%%%%%%%%%%%%%%%%%%%%%%%%%%%%%%%%%%%%%%%%%%%%
\end{document}